\documentclass[aps,pra,showpacs,
amssymb,twocolumn
]{revtex4} 
\usepackage{graphicx}
\usepackage{amsmath}
\usepackage{amsfonts}
\usepackage{amssymb}
\usepackage{epsfig,libertine}
\usepackage[libertine]{newtxmath}
\usepackage{color}
\usepackage{dsfont, hyperref, xcolor}
\usepackage{comment}
\usepackage{enumerate}

\usepackage{amsthm} 
\newcommand{\Tr}[1]{\text{Tr}\left\{#1\right\}}
\newcommand{\ParTr}[2]{\text{Tr}_{#1}\left\{#2\right\}}
\newcommand{\bra}[1]{\langle#1\vert}
\newcommand{\ket}[1]{\vert#1\rangle}

\begin{document}

\title{Resource theories with completely passive states and passive states as free states}

\author{Gianluca~Francica}
\address{Dipartimento di Fisica e Astronomia e Sezione INFN, Università di Padova, via Marzolo 8, 35131 Padova, Italy}

\date{\today}

\begin{abstract}
{\color{black}Work extractable from quantum system can be viewed as a resource to perform some tasks. Here, we try to related the work extractable to some resource theory. To do this, we consider completely passive states and passive states as free states and we formulate resource theories where the maximum work extractable is a monotone, by identifying the free operations. In particular, we study how the free operations act on the free states,  showing how a sort of temperature plays a pivotal role in these resource theories. }
\end{abstract}


\maketitle

\section{Introduction}
Work extraction processes are of fundamental importance in quantum thermodynamics~\cite{bookthermo}, for instance recently they played a certain role in the context of quantum batteries~\cite{Campaioli23}. For finite thermally isolated quantum systems, the maximum work extractable through a cycle is upper bounded by a free energy that is obtained by minimizing the final energy with the constraint that the von Neumann entropy remains constant~\cite{Bera19}. In general, the maximum work is equal to the quantum ergotropy~\cite{Allahverdyan04} and the states with zero ergotropy are passive. 
The free energy can be extracted in the limit of an infinite number of copies~\cite{Alicki13} and the states that are passive for any number of copies are completely passive.
In this paper, the maximum work extractable is viewed as a resource, and thus linked to the paradigm of the resource theories (see, e.g., Ref.~\cite{Goold16} for the role of resource theories in quantum thermodynamics).
For instance, a paradigmatic resource theory views the entanglement as a resource~\cite{Horodecki09}.
In this resource theory, states without entaglement are separable states, which are viewed as free states. Local operations and classical communication do not generate entanglement, leaving separable states separable, and thus are viewed as free operations.
Similarly, other resource theories can be formulated, for instance a resource theory of quantum coherence~\cite{Streltsov17}. 
Furthermore, a resource theory of quantum thermodynamics has been introduced in Ref.~\cite{Janzing00}, where the free operations are thermal operations and do not increase the nonequilibrium Helmholtz free energy (see, e.g., Ref.~\cite{Lostaglio19}). From a microscopic point of view, work can be extracted through thermal operations only when the system interacts with an environment at a certain initial temperature.
In contrast, we focus on reversible work extractions, discussing two resource theories where the resources are the free energy and the ergotropy, respectively. {\color{black} Our main motivation is to understand what is the general structure of the processes that can be performed without generating free energy or ergotropy so as to use these resources.}

Thus, after introducing some preliminary notions in Sec.~\ref{sec.preli}, we show how the free energy and the ergotropy can be viewed as distance-based monotone of the corresponding resource theories in Secs.~\ref{sec.cpfree} and~\ref{sec.pfree}, respectively. A unified picture follows for the two resource theories by expressing the monotones in terms of a unique distance (in broad sense). In particular, the distance is achieved by multiplying the relative entropy by a sort of nonequilibrium temperature. Then, free operations can increase the nonequilibrium temperature of the free states at most by a certain factor.
We examine the monotonicity properties of the free energy and the ergotropy, the contractivity of the distance and other properties, also with the help of specific free operations.
{\color{black}We also introduce some other monotones for the resource theories in Sec.~\ref{sec.monotones}.}
Finally, we summarize and discuss further the results achieved in Sec.~\ref{sec.conclusions}.

\section{Preliminaries}\label{sec.preli}
We start our discussion by introducing some preliminary notions, which are some rudiments about resource theories (see Sec.~\ref{sec.resources}) and the work extraction (see Sec.~\ref{sec.ergotropy}).

\subsection{Resource theories}\label{sec.resources}
In general, a resource theory can be defined from a set of free states $\mathcal S$. Free operations are some completely positive and trace preserving (CPTP) maps $\Lambda(\rho) = \sum_i K_i \rho K^\dagger_i$, such that $\Lambda(\mathcal S)\subseteq \mathcal S$, where the Kraus operators $K_i$ are such that $\sum_i K^\dagger_i K_i =  I$ and $I$ is the identity matrix. In general, the free operations form a set $\mathcal O \subseteq \mathcal M =\{\Lambda | \Lambda(\mathcal S)\subseteq \mathcal S \}$, which can be a subset of $\mathcal M$ since they can satisfy additional constraints.
The resource is quantified by a monotone, which is a functional $M(\rho)$ satisfying two minimal properties, which are (i) the nonnegativity property  $M(\rho)\geq 0$ and $M(\rho)=0$ if and only if $\rho \in \mathcal S$ and (ii) the monotonicity property  $M(\Lambda(\rho))\leq M(\rho)$ for all $\Lambda \in \mathcal O$.
A monotone can also satisfy (iii) the strong monotonicity property  $\sum_i q_i M(\sigma_i) \leq M(\rho)$, where $\sigma_i=K_i \rho K_i^\dagger/q_i$, with $q_i=\Tr{K^\dagger_iK_i \rho}$. Furthermore, a monotone can be (iv) convex, i.e., $M(\sum_i p_i \rho_i) \leq \sum_i p_i M(\rho_i)$ for any convex combination of density matrices $\sum_i p_i \rho_i$. Typically, we get a measure only if the monotone is (v) additive under tensor products, i.e., $M(\rho_1 \otimes \rho_2) = M(\rho_1) + M(\rho_2)$.
A distance-based monotone is defined  as
\begin{equation}
M(\rho) = \inf_{\sigma\in \mathcal S} D(\rho, \sigma)
\end{equation}
where $D(\rho, \sigma)$ is a distance. {\color{black}Here ``distance'' is understood in a rather broad term}, i.e., $D(\rho, \sigma)$ is a nonnegative functional such that $D(\rho,\sigma)=0$ if and only if $\rho=\sigma$, so that the monotone satisfies (i). 
For instance, the distance can be chosen to be the quantum relative entropy $S(\rho||\sigma)=\Tr{\rho(\ln \rho-\ln \sigma)}$, so that it is contractive, $D(\Lambda(\rho),\Lambda(\sigma))\leq D(\rho,\sigma)$ for any CPTP map, and the monotone satisfies (ii).

We can formulate different resource theories depending on what the resource is.
For instance, the quantum entanglement can be viewed as a resource, such that the free state are separable states and the free operations are obtained through local operations and classical communication~\cite{Horodecki09}. When the quantum coherence is viewed as a resource, the incoherent states, i.e., the states diagonal with respect to a fixed basis, are free states, and the largest set of free operations is formed by the maximally incoherent operations, which are defined such that they map incoherent states into incoherent states~\cite{Streltsov17}.
In particular, we are going to follow this approach in the next sections. In the resource theory of quantum thermodynamics~\cite{Lostaglio19}, the free operations are the so-called thermal operations, which are defined as $\Lambda_{th}(\rho)=\ParTr{E}{U \rho\otimes \gamma^E_\beta U^\dagger}$ with unitary operators $U$ that commutate with the total Hamiltonian.
In detail, the thermal operations act on the Hilbert space of a system $S$ having Hamiltonian $H$, and are defined by performing a Stinespring dilation with an environment $E$ having Hamiltonian $H_E$ and prepared in the Gibbs state $\gamma^E_\beta=e^{-\beta H_E}/Z_E$ at the temperature $T=\beta^{-1}$, with $Z_E=\Tr{e^{-\beta H_E}}$, 
and with a global unitary operator $U$ such that $[U,H+H_E]=0$.
%
The thermal operations preserve the Gibbs state of the system $S$, $\Lambda_{th}(\gamma_\beta)=\gamma_\beta$, thus the Gibbs state  $\gamma_\beta=e^{-\beta H}/Z$, with $Z=\Tr{e^{-\beta H}}$, is the free state. It results that not all the CPTP maps that preserve the Gibbs state $\gamma_\beta$, forming the set of Gibbs-preserving operations, are thermal operations (see, e.g., Ref.~\cite{Faist15}).
Furthermore, the nonequilibrium Helmholtz free energy defined as $F_\beta(\rho) = T S(\rho||\gamma_\beta)- T \ln Z$ does not increase under the action of the thermal operations, i.e., $F_\beta(\Lambda_{th}(\rho))\leq F_\beta(\rho)$, so that it is a monotone.
The maximum work extractable is given by the difference  $\Delta  F_\beta(\rho) = F_\beta(\rho)- F_\beta(\gamma_\beta)=T S(\rho||\gamma_\beta)$.
We note that in this case a work extraction is microscopically performed through an environment $E$, which initially is in the equilibrium state $\gamma^E_\beta$.
In contrast, in the following we focus on a reversible work extraction, such that the system $S$ is thermally isolated. Thus, in order to extract a nonzero average work, the unitary time-evolution operator must not commutate with the Hamiltonian.
We aim to define some free operations that do not increase the maximum work extractable, which is viewed as a resource.
In particular, we find that in the definition of these free operations some sort of temperature plays a pivotal role.

\subsection{Work extraction}\label{sec.ergotropy}
Let us introduce the resources that we aim to investigate from the resource theories perspective. Following Ref.~\cite{Allahverdyan04}, we consider a finite quantum system $S$ 
having a  Hilbert space $\mathcal H$ with dimension $d$. The Hamiltonian is expressed as
\begin{equation}\label{eq.hami}
H=\sum_{k} \epsilon_k \ket{\epsilon_k}\bra{\epsilon_k}\,,
\end{equation}
with $\epsilon_{k}< \epsilon_{k+1}$, and the initial state is given by
\begin{equation}
\rho = \sum_k r_k \ket{r_k} \bra{r_k}\,,
\end{equation}
with $r_{k}\geq r_{k+1}$. In order to extract work from the system $S$, we perform a unitary cycle, i.e., a cyclical change of some Hamiltonian parameters, so that the time-evolution from the initial time $t=0$ to the final time $t=\tau$ is generated by a time-dependent Hamiltonian $H(t)$ such that $H(0)=H(\tau)=H$. The resulting time-evolution operator is $U=\mathcal T e^{-i\int_0^\tau H(t)dt}$, where $\mathcal T$ is the time-ordering operator, and the final state at the time $\tau$ is $\rho'=U\rho U^\dagger$. Since the system is thermally isolated, from the first law of the thermodynamics the average work extracted in the unitary cycle is
\begin{equation}\label{eq.workave}
W(\rho,U)=E(\rho)-E(U \rho U^\dagger)\,,
\end{equation}
where $E(\rho)=\Tr{H\rho}$ is the average energy of the state $\rho$, which is calculated with respect to the Hamiltonian $H$.
We note that the von Neumann entropy, defined as $S(\rho)=-\Tr{\rho\ln\rho}$, remains constant during the cycle, i.e., $S(\rho')=S(\rho)$, then by minimizing the energy $E(\rho')$ over the final states $\rho'$ having an entropy equal to its initial value, 
we find the upper bound $W(\rho,U)\leq F(\rho)$ for the average work, where
\begin{equation}
F(\rho) = E(\rho) - E(\gamma_\rho)\geq 0\,,
\end{equation}
with $\gamma_\rho = \gamma_{\beta(\rho)}$ and $\beta(\rho)$ solution of a nonlinear equation for $\beta$, which is $S(\rho_\beta)=S(\rho)$. Here, we simply refer to $F(\rho)$ as free energy, as in Ref.~\cite{Bera19}.
In general, 
the maximum work extractable is achieved by maximizing  Eq.~\eqref{eq.workave} over all the unitary cycles and it is equal to the ergotropy~\cite{Allahverdyan04}, here denoted with $\mathcal E (\rho)$, i.e.,
\begin{equation}\label{eq.ergotropy def}
\mathcal E (\rho) = \max_{U} W(\rho,U) \geq 0\,.
\end{equation}
The ergotropy reads
\begin{equation}
\mathcal E(\rho) = E(\rho) - E(P_\rho)\,,
\end{equation}
where $P_\rho= \sum_k r_k \ket{\epsilon_k}\bra{\epsilon_k}$ is a passive state. In particular, the maximum average work $\mathcal E (\rho)$ is obtained for a unitary operator $U_\rho = \sum_k e^{i\phi_k} \ket{\epsilon_k}\bra{r_k}$, where $\phi_k$ are some arbitrary real phases, so that the final state $\rho'=U_\rho \rho U_\rho^\dagger$ is the passive state $P_\rho$.
In particular, we note that $P_{U\rho U^\dagger}=P_\rho$ for any unitary operator $U$. In general, the set  $\Pi$  of the passive states is formed by all the states $P$ having zero ergotropy, i.e., such that $\mathcal E(P)=0$.  In other words, the ergotropy $\mathcal E(\rho)$ is zero if and only if the state $\rho$ is passive. 
{\color{black}All the Gibbs states $\gamma_\beta$ with $\beta\geq 0$ are passive states}, and in general the set $\Pi$ is formed by all the states $P=\sum_k p_k \ket{\epsilon_k}\bra{\epsilon_k}$ with $p_k\geq p_{k+1}$. 
{\color{black}If the energy spectrum is degenerate, i.e., $\epsilon_k\leq \epsilon_{k+1}$, the set $\Pi$ is formed by all the states $P=\sum_k p_k U\ket{\epsilon_k}\bra{\epsilon_k}U^\dagger$ with $p_k\geq p_{k+1}$ and $U$ arbitrary unitary operator such that $[U,H]=0$. In particular, one can check that all the results that we are going to derive trivially hold also for $\epsilon_k\leq \epsilon_{k+1}$. We focused on $\epsilon_k < \epsilon_{k+1}$ only to introduce the passive state $P_\rho$, which is well defined even if there is some degeneracy in the energy.
}

Let us consider an ensemble of $n$ copies of the system $S$, as in Ref.~\cite{Alicki13}, where any copy has the Hamiltonian $H$ and is prepared in the state $\rho$. The state $\rho$ is  completely passive if the state $\rho^{\otimes n}$ is passive with respect to the total Hamiltonian for any $n$. In general, the set $C$ of the completely passive states is formed by all the Gibbs states $\gamma_\beta$ with $\beta\geq 0$.
This can be understood by noting that $\mathcal E ( \rho^{\otimes n}) \leq n F(\rho)$ for any $n$. 
Then, by noting that $F(\rho)=0$ if and only if $\rho=\gamma_\beta$ with $\beta\geq 0$, we get $\mathcal E ( \rho^{\otimes n})=0$ for any $n$ if $\rho$ is a Gibbs state.
Furthermore, it is possible to perform a global unitary cycle such that~\cite{Alicki13,Francica24}
\begin{equation}\label{eq.Alicki-Fannes}
\mathcal E ( \rho^{\otimes n}) \sim n F(\rho)
\end{equation}
as $n\to \infty$. Then, if $\rho\neq \gamma_\beta$ with $\beta\geq 0$, there exists a sufficient large integer $n$ such that
$\mathcal E ( \rho^{\otimes n}) >0$. Thus, we deduce that  $\mathcal E ( \rho^{\otimes n})=0$ for any $n$ if and only if $\rho$ is a Gibbs state.
In the end, we note that, since $\gamma_{U\rho U^\dagger}=\gamma_\rho$ for any unitary operator $U$, the free energy and the ergotropy are related by the equation
\begin{equation}\label{eq.ergoandfree}
\mathcal E(\rho) = F(\rho)-F(P_\rho)\,.
\end{equation}
\begin{proof}
To prove it, we note that
\begin{equation}
\mathcal E(\rho)= E(\rho)-E(P_\rho) = E(\rho)-E(\gamma_\rho)-E(P_\rho)+E(\gamma_{P_\rho})\,,
\end{equation}
from which it follows Eq.~\eqref{eq.ergoandfree}.
\end{proof}

\section{Completely passive states as free states}\label{sec.cpfree}
We start to consider the set $C$ of the completely passive states, $C=\{\gamma_\beta | \beta\geq 0\}$, as the set of the free states. 
The set $\mathcal O_{cp}$ of the free operations is formed by all the CPTP maps $\Lambda$ that satisfy (F,i) $\Lambda(C)\subseteq C$ 
and (F,ii) $F(\Lambda(\rho))\leq F(\rho)$ for all $\rho$.
In particular, the condition (F,ii) implies (F,i).
\begin{proof}
If $F(\Lambda(\rho))\leq F(\rho)$ for all $\rho$, then $F(\Lambda(\gamma))\leq F(\gamma)=0$, and thus $\Lambda(\gamma) \in C$ for all $\gamma\in C$.
\end{proof}
Then,  $\mathcal O_{cp}\subseteq \mathcal M_{cp}=\{\Lambda|\Lambda(C)\subseteq C\}$ and  $M_{cp}(\rho)=\min_{\gamma\in C} S(\rho||\gamma)$ is a monotone, since $M_{cp}(\Lambda(\rho))\leq M_{cp}(\rho)$ for all $\Lambda\in\mathcal M_{cp}$.
In contrast, by defining the temperature functional $T(\gamma)$ such that $T(\gamma_\beta)=\beta^{-1}$, the free energy $F(\rho)$ can be expressed as~\cite{Bera19}
\begin{equation}\label{eq.F}
F(\rho)= \min_{\gamma\in C} T(\gamma) S(\rho||\gamma)\,,
\end{equation}
and due to the temperature $T(\gamma)$ that multiplies the relative entropy $S(\rho||\gamma)$, the condition (F,ii) is not satisfied for all the maps belonging to $\mathcal M_{cp}$. We find that the condition (F,ii) is satisfied if and only if the condition (F,i) is satisfied and
\begin{equation}\label{eq.F1}
\eta_{\Lambda}(\gamma) T(\Lambda(\gamma)) \leq T(\gamma)\,, \forall \gamma \in C\,,
\end{equation}
where the condition (F,i) implies
\begin{equation}\label{eq.F2}
\eta_{\Lambda}(\gamma) \leq \max_\rho \frac{S(\Lambda(\rho)||\Lambda(\gamma))}{S(\rho||\gamma)}\leq 1\,,\forall \gamma\in C\,,
\end{equation}
and we defined the contraction factor
\begin{equation}\label{eq.etaF}
\eta_{\Lambda}(\gamma) = \frac{1}{T(\Lambda(\gamma))} \max_\rho \frac{F(\Lambda(\rho))}{S(\rho||\gamma)}\,.
\end{equation}
\begin{proof}
From Eq.~\eqref{eq.F}, we get $F(\rho)\leq T(\gamma) S(\rho||\gamma)$ for all $\gamma\in C$. Then, the condition (F,ii), which is equivalent to \begin{equation}
\max_\rho \frac{F(\Lambda(\rho))}{F(\rho)}\leq 1\,,
\end{equation}
is satisfied if and only if
\begin{equation}
\max_\rho \frac{F(\Lambda(\rho))}{S(\rho||\gamma)} \leq T(\gamma)\,,\forall \gamma \in C\,,
\end{equation}
from which we get Eq.~\eqref{eq.F1} by considering the definition for $\eta_\Lambda(\gamma)$ given in Eq.~\eqref{eq.etaF}. On the other hand, the condition (F,i) implies that $\Lambda(\gamma)\in C$, then $F(\Lambda(\rho))\leq T(\Lambda(\gamma))S(\Lambda(\rho)||\Lambda(\gamma))$, so that
\begin{equation}
\max_\rho \frac{F(\Lambda(\rho))}{S(\rho||\gamma)} \leq T(\Lambda(\gamma)) \max_\rho \frac{S(\Lambda(\rho)||\Lambda(\gamma))}{S(\rho||\gamma)}\leq T(\Lambda(\gamma))\,,
\end{equation}
from which we get Eq.~\eqref{eq.F2}.
\end{proof}
Thus, the set of free operations $\mathcal O_{cp}$ is formed by all the CPTP maps $\Lambda$ that satisfy the conditions (F,i) and (F,iii), where (F,iii) is the condition in Eq.~\eqref{eq.F1}.
We can see that the condition (F,i) does not imply (F,ii) by considering a map $\Lambda_\beta$ such that $\Lambda_\beta(\gamma)=\gamma_\beta$. Similarly to the maps considered in Refs.~\cite{Faist15,Tajima24}, the map $\Lambda_\beta$  can be realized as
\begin{equation}\label{eq.example}
\Lambda_\beta(\rho) = \Tr{(I-\ket{\psi_d}\bra{\psi_d})\rho} \sigma + \bra{\psi_d}\rho \ket{\psi_d} \sigma'\,,
\end{equation}
where  $\ket{\psi_d} = \sum_k \ket{\epsilon_k}/\sqrt{d}$, $\sigma=d/(d-1)( \gamma_\beta - \sigma'/d)$ and $\sigma'$ is any density matrix such that $\sigma\geq 0$.
The map in Eq.~\eqref{eq.example} is such that $\Lambda_\beta (\rho) = \gamma_\beta$ for any incoherent state $\rho$, i.e.,
for any state diagonal with respect to the energy basis. 
Then, $\Lambda_\beta(\gamma_{\beta'})=\gamma_\beta$ for all $\beta'$ and thus $\Lambda_\beta \in \mathcal M_{cp}$, so that the condition (F,i) is satisfied.
Since, the image of the map $\Lambda_\beta$ is not a subset of $C$, we get $\eta_\Lambda(\gamma)>0$ and Eq.~\eqref{eq.F1} is satisfied only if we choose the parameter $\beta$ of the map such that $\beta^{-1}\leq  T(\gamma) / \eta_{\Lambda}(\gamma) $ when $\rho=\gamma\in C$, and thus the condition (F,ii) is not always satisfied.

Therefore, the free operations $\Lambda \in \mathcal O_{cp}$ can increase the temperature of the free states $\gamma$ at most by a certain factor $1/ \eta_{\Lambda}(\gamma)\geq 1$. If the free operation $\Lambda\in \mathcal O_{cp}$ is such that $\eta_{\Lambda}(C)=1$, then it does not increase the  temperature of the free states, i.e., $T(\Lambda(\gamma))\leq T(\gamma)$ for all $\gamma\in C$.
We define a distance $D(\rho,\sigma)$ such that, for $\sigma=\gamma\in C$ and $\gamma$ different from the ground-state, i.e., $\gamma\neq \ket{\epsilon_1}\bra{\epsilon_1}$, it reads
\begin{equation}\label{eq.distF}
D(\rho,\gamma) = T(\gamma) S(\rho||\gamma)\,,
\end{equation}
so that $D(\rho,\gamma)=0$ if and only if $\rho=\gamma$. Then, from Eq.~\eqref{eq.F} we deduce that the free energy is a based-distance monotone, since it can be expressed as
\begin{equation}
F(\rho)= \min_{\gamma\in C} D(\rho,\gamma)\,.
\end{equation}
Given a free operation $\Lambda\in \mathcal O_{cp}$, by noting that the relative entropy $S(\rho||\gamma)$ is contractive, from Eq.~\eqref{eq.F1} it follows
\begin{equation}
D(\Lambda(\rho),\Lambda(\gamma))= T(\Lambda(\gamma)) S(\Lambda(\rho)||\Lambda(\gamma))\leq D(\rho,\gamma)/\eta_\Lambda(\gamma)\,,
\end{equation}
so that the distance in Eq.~\eqref{eq.distF} is contractive  under the action of free operations $\Lambda\in \mathcal O_{cp}$ such that $\eta_{\Lambda}(C)=1$. In general, for $\Lambda\in \mathcal O_{cp}$ we get $D(\Lambda(\rho),\Lambda(\gamma))\leq D(\rho,\gamma)$ if and only if the bound in Eq.~\eqref{eq.F2} is saturated.
\begin{proof}
We note that for $\Lambda\in\mathcal O_{cp}$, from Eq.~\eqref{eq.F1}
\begin{equation}
\eta_{\Lambda}(\gamma) = \max_\rho \frac{S(\Lambda(\rho)||\Lambda(\gamma))}{S(\rho||\gamma)}\,,\forall \gamma\in C\,,
\end{equation}
if and  only if
\begin{equation}
 \max_\rho \frac{S(\Lambda(\rho)||\Lambda(\gamma))}{S(\rho||\gamma)}\leq \frac{T(\gamma)}{T(\Lambda(\gamma))}\,,\forall \gamma\in C\,,
\end{equation}
i.e., $D(\Lambda(\rho),\Lambda(\gamma))\leq D(\rho,\gamma)$ for all $\rho$ and for all $\gamma \in C$.
\end{proof}

Furthermore, we note that $F(\rho)$  satisfies the strong monotonicity property, i.e.,
\begin{equation}\label{eq.stronFdef}
\sum_i q_i F(\sigma_i) \leq F(\rho)\,,
\end{equation}
for any free operation $\Lambda\in \mathcal O_{cp}$ such that $\gamma_i\in C$ and
\begin{equation}\label{eq.strongFcond}
T(\gamma)S(\rho||\gamma)\geq \sum_i q_i T(\gamma_i)S(\sigma_i||\gamma_i)
\end{equation}
for all $\gamma\in C$, where $\sigma_i=K_i \rho K_i^\dagger/q_i$, $q_i=\Tr{K^\dagger_iK_i \rho}$, $\gamma_i=K_i \gamma K_i^\dagger/\Tr{K^\dagger_i K_i \gamma}$ and $\Lambda(\rho)=\sum_i K_i \rho K^\dagger_i=\sum_i q_i \sigma_i$. For instance, Eq.~\eqref{eq.strongFcond} is satisfied if $T(\gamma_i)\leq T(\gamma)$ for all $i$, since in general $S(\rho||\gamma)\geq \sum_i q_i S(\sigma_i||\gamma_i)$ (see, e.g., Ref.~\cite{Vedral98}).
\begin{proof}
To prove it, we note that
\begin{equation}
F(\rho)=T(\gamma_\rho) S(\rho||\gamma_\rho)\,,
\end{equation}
thus from Eq.~\eqref{eq.strongFcond}, since $\gamma_i\in C$ for all $i$, we get
\begin{equation}
F(\rho)\geq \sum_i q_i T(\gamma_i)S(\sigma_i||\gamma_i)\geq \sum_i q_i T(\gamma_{\sigma_i})S(\sigma_i||\gamma_{\sigma_i})\,,
\end{equation}
from which it follows Eq.~\eqref{eq.stronFdef}.
\end{proof}
We note that the set $C$ is not convex for $d>2$, i.e., the convex combination of two Gibbs states is not always a Gibbs state, and the free energy $F(\rho)$ is not convex, e.g., for $\rho=\sum_i p_i \gamma_i \notin C$, we get $F(\rho)>\sum_i p_iF(\gamma_i)=0$ if $\gamma_i\in C$ for all $i$.
Then, a free operation $\Lambda$ that satisfies the strong monotonicity property is such that $K_i \gamma K_i^\dagger/\Tr{K^\dagger_i K_i \gamma}=\Lambda(\gamma)$ for all $i$.
Furthermore, if we consider two systems with Hamiltonian $H_1$ and $H_2$, and the composite system with Hamiltonian $H=H_1+H_2$,  we recall that the free energy is super-additive under tensor products~\cite{Bera19}, i.e., $F(\rho_1\otimes \rho_2) \geq F(\rho_1) + F(\rho_2)$, and it is additive if $\beta(\rho_1)=\beta(\rho_2)$, thus it cannot be considered as a measure.

The set $\mathcal O_{cp}$ is formed by all the maps $\Lambda$ such that $\Lambda(\gamma)=\gamma_{\beta}$ for all $\gamma\in C$ with $\beta\geq \eta_{\Lambda}(\gamma)\beta(\gamma)$, where $\beta(\gamma)=1/T(\gamma)$. Let us give some examples of these free operations.
The dephasing map $\Delta(\rho) = \sum_k \ket{\epsilon_k}\bra{\epsilon_k} \rho \ket{\epsilon_k}\bra{\epsilon_k}$ belongs to $\mathcal O_{cp}$ since $\Delta(\gamma_\beta)=\gamma_\beta$. Analogously, the partial dephasing map
\begin{equation}
\Gamma(\rho) = \sum_k \ket{\epsilon_k}\bra{\epsilon_k} \rho \ket{\epsilon_k}\bra{\epsilon_k} + \sum_{k\neq l} \alpha_{kl}\ket{\epsilon_k}\bra{\epsilon_k} \rho \ket{\epsilon_l}\bra{\epsilon_l}
\end{equation}
with $|\alpha_{kl}|\leq 1$ belongs to $\mathcal O_{cp}$. In particular, although these maps preserve the temperature of all the  free states, they can decrease the free energy.
To give an example of a free operation that does not preserve the Gibbs state, we consider the map $\Lambda_\beta$ in Eq.~\eqref{eq.example}.
We get $\Lambda_{\tilde{\beta}(\rho)}\in \mathcal O_{cp}$ where $\tilde \beta(\rho)$ is any function such that $\tilde \beta(\rho)\geq \beta(\rho)$ when $\rho\in C$. In this case, the map is not linear since the Kraus operators depend on $\rho$ through $\tilde \beta(\rho)$.
We note that the distance $D(\rho,\gamma)$ in Eq.~\eqref{eq.distF} is contractive under the action of all these maps.
For the limit case of a thermalizing map $\mathcal T_\beta$ such that $\mathcal T_\beta(\rho)=\gamma_\beta$ for all $\rho$ (e.g., achieved from the map $\Lambda_\beta$ for $\sigma'=\gamma_\beta$), we get $\eta_{\mathcal T_\beta}(C)=0$, thus the distance $D(\rho,\gamma)$ is not contractive under the action of the map $\mathcal T_\beta$. In this case, the  temperature of the free states can be arbitrarily increased, i.e., $\mathcal T_\beta \in \mathcal O_{cp}$ for any $\beta\geq 0$.
Similarly, the thermal operations $\Lambda_{th|\beta}$ for a given value of $\beta$, or more generally all the operations that preserve the Gibbs state $\gamma_\beta$ for a certain value of $\beta$, belong to $\mathcal O_{cp}$ if the value of $\beta$ depends on $\rho$, e.g., it  is $\beta=\beta(\rho)$ when $\rho\in C$. Of course, these maps can be not linear since $\beta$ depends on $\rho$.
%
%
We note that  $M_{cp}(\rho)= S(\rho||\gamma_{\hat\beta(\rho)})$, where $\hat\beta(\rho)$ minimizes the relative entropy $S(\rho||\gamma_{\beta})$ over $\beta\geq 0$, thus $M_{cp}(\rho)$ is related to the nonequilibrium Helmholtz free energy as $M_{cp}(\rho) = \hat\beta(\rho) \Delta F_{\hat\beta(\rho)}(\rho) $.

\section{Passive states as free states}\label{sec.pfree}
Let us try to extend the results achieved to the case of free states that are passive states.
We consider the set $\Pi$ of the passive states 
 as the set of the free states.
The set $\mathcal O_{p}$ of the free operations is formed by all the CPTP maps $\Lambda$ that satisfy (E,i) $\Lambda(\Pi)\subseteq \Pi$ 
and (E,ii) $\mathcal E(\Lambda(\rho))\leq \mathcal E(\rho)$ for all $\rho$.
Similarly to the free energy case, the condition (E,ii) implies (E,i).
\begin{proof}
If $\mathcal E(\Lambda(\rho))\leq \mathcal E(\rho)$ for all $\rho$, then $\mathcal E(\Lambda(P))\leq \mathcal E(P)=0$, and thus $\Lambda(P) \in \Pi$ for all $P\in \Pi$.
\end{proof}
Then, $\mathcal O_{p}\subseteq \mathcal M_{p}=\{\Lambda|\Lambda(\Pi)\subseteq\Pi\}$ and  $M_{p}(\rho)=\min_{P\in \Pi} S(\rho||P)$ is a monotone, since $M_{p}(\Lambda(\rho))\leq M_{p}(\rho)$ for all $\Lambda\in\mathcal M_{p}$.
We note that the ergotropy is equal to the free energy, $\mathcal E(\rho)=F(\rho)$, when $P_\rho$ is a completely passive state, i.e., $P_\rho=\gamma_\rho$.
Then, to generalize the results achieved for the free energy to the case of the passive states as free states, we try to define {\color{black}something like} a nonequilibrium temperature for the passive state $P$ such that it is equal to the temperature functional $T(P)$ when $P=\gamma\in C$. Thus, from the monotone $M_{p}(\rho)$, similarly to Eq.~\eqref{eq.F}, we define a functional
\begin{equation}\label{eq.phi}
\Phi(\rho)= \min_{P\in \Pi} T(P|\rho) S(\rho||P)\,,
\end{equation}
where $T(P|\rho)$ is some functional that plays the role of nonequilibrium temperature for the passive state $P$.
Of course there are infinite ways to define a functional $T(P)$ of a passive state $P$ such that $T(P)=\beta^{-1}$ when $P=\gamma_\beta$.
We refer to this functional $T(P)$ as nonequilibrium temperature since it gives the usual temperature of $P$ when $P$ is a Gibbs state.
When we perform the minimization in Eq.~\eqref{eq.phi}, we can select a certain definition of the nonequilibrium temperature, among all the possible definitions, depending on the state $\rho$, achieving in this way a nonequilibrium temperature $T(P|\rho)$ that depends on $\rho$.
The minimum in Eq.~\eqref{eq.phi} can be achieved for a unique passive state $P^*$, such that $\Phi(\rho)= T(P^*|\rho) S(\rho||P^*)$.
Given a state $\rho$, we can choose the nonequilibrium temperature $T(P|\rho)$  such that $P^*=P_\rho$ and the functional in Eq.~\eqref{eq.phi} is equal to the ergotropy, i.e., $\Phi(\rho)=\mathcal E(\rho)$. In detail, we find that $P^*=P_\rho$ and
\begin{equation}\label{eq.ergodist}
\mathcal E(\rho) = \min_{P\in \Pi} T(P|\rho) S(\rho||P)
\end{equation}
for the nonequilibrium temperature
\begin{equation}\label{eq.Tnoneq}
T(P|\rho)  = \frac{\mathcal E(\rho)}{\Tr{(P_\rho-\rho)\ln P}}\,, 
\end{equation}
which is such that $T(\gamma_\beta|\rho)=T(\gamma_\beta)=\beta^{-1}$ for any $\rho$.
\begin{proof}
To prove that $P^*=P_\rho$, we consider that the first variation of $T(\rho|P)S(\rho||P)$ with respect to $P$ is zero, i.e.,
\begin{equation}\label{eq.varia}
\delta \left( T(P|\rho) S(\rho||P)+\lambda \Tr{P} \right)=0\,,
\end{equation}
for $P=P^*$, where $\lambda$ is a Lagrange multiplier. Eq.~\eqref{eq.varia} is satisfied for any $\delta P$ if
\begin{eqnarray}\nonumber
&&\frac{\mathcal E(\rho)}{\Tr{(P_\rho-\rho)\ln P}^2} (S(\rho)+\Tr{\rho \ln P})(P_\rho -\rho) \\
&&- \frac{\mathcal E(\rho)}{\Tr{(P_\rho-\rho)\ln P}} \rho + \lambda P =0\,,
\end{eqnarray}
which is satisfied if $P=P_\rho$ and $\lambda = \frac{\mathcal E(\rho)}{\Tr{(P_\rho-\rho)\ln P_\rho}}$ (it is enough to consider that $S(\rho)=S(P_\rho)$ and thus $S(\rho)+\Tr{\rho \ln P}= \Tr{(\rho-P_\rho)\ln P_\rho}$).
\end{proof}
In particular, we note that, from Eq.~\eqref{eq.ergodist}, it follows  the inequality
\begin{equation}\label{eq.ine}
T(P|\rho) S(\rho||P) \geq \mathcal E(\rho)\,,\,\forall P\in \Pi\,,\forall \rho,
\end{equation}
where the equality holds if and only if $P=P_\rho$.
Similarly to the free energy case, the condition (E,ii) is not satisfied for all the maps belonging to $\mathcal M_{p}$. Since the selection of the nonequilibrium temperature depends on $\rho$, we find that the condition (E,ii) is satisfied if and only if the condition (E,i) is satisfied and
\begin{equation}\label{eq.E1}
\eta_{\Lambda}(P|\rho) T(\Lambda(P)|\Lambda(\rho)) \leq T(P|\rho)\,, \forall P \in \Pi\,,\forall \rho\,,
\end{equation}
where the condition (E,i) implies
\begin{equation}\label{eq.E2}
\eta_{\Lambda}(P|\rho) \leq \frac{S(\Lambda(\rho)||\Lambda(P))}{S(\rho||P)}\leq 1\,,\forall P\in \Pi\,,\forall \rho\,,
\end{equation}
and we defined the contraction factor
\begin{equation}\label{eq.etaergo}
\eta_{\Lambda}(P|\rho) = \frac{1}{T(\Lambda(P)|\Lambda(\rho))}  \frac{\mathcal E(\Lambda(\rho))}{S(\rho||\gamma)}\,.
\end{equation}
\begin{proof}
From Eq.~\eqref{eq.ine}, the condition (E,ii), $\mathcal E(\Lambda(\rho))\leq \mathcal E(\rho)$ for all $\rho$, is satisfied if and only if
\begin{equation}
\frac{\mathcal E(\Lambda(\rho))}{S(\rho||P)} \leq T(P|\rho)\,,\forall P \in \Pi\,,\forall \rho\,,
\end{equation}
from which we get Eq.~\eqref{eq.E1} by considering the definition for $\eta_{\Lambda}(P|\rho)$ in Eq.~\eqref{eq.etaergo}. On the other hand, the condition (E,i) implies that $\Lambda(P)\in \Pi$, then $\mathcal E(\Lambda(\rho))\leq T(\Lambda(P)|\Lambda(\rho))S(\Lambda(\rho)||\Lambda(P))$, so that from Eq.~\eqref{eq.etaergo} we get Eq.~\eqref{eq.E2}.
\end{proof}
Thus, the set of free operations $\mathcal O_{p}$ is formed by all the CPTP maps $\Lambda$ that satisfy the conditions (E,i) and (E,iii), where (E,iii) is the condition in Eq.~\eqref{eq.E1}.
Similarly to the free energy case, we note that $\mathcal E(\rho)$ satisfies the strong monotonicity property, i.e.,
\begin{equation}\label{eq.stronEdef}
\sum_i q_i \mathcal E(\sigma_i) \leq \mathcal E(\rho)\,,
\end{equation}
for any free operation $\Lambda\in \mathcal O_{p}$ such that $P_i\in \Pi$ and
\begin{equation}\label{eq.strongEcond}
T(P|\rho)S(\rho||P)\geq \sum_i q_i T(P_i|\sigma_i)S(\sigma_i||P_i)
\end{equation}
for all $P\in \Pi$, where $\sigma_i=K_i \rho K_i^\dagger/q_i$, $q_i=\Tr{K^\dagger_iK_i \rho}$, $P_i=K_i P K_i^\dagger/\Tr{K^\dagger_i K_i P}$ and $\Lambda(\rho)=\sum_i K_i \rho K^\dagger_i=\sum_i q_i \sigma_i$. For instance, Eq.~\eqref{eq.strongEcond} is satisfied if $T(P_i|\sigma_i)\leq T(P|\rho)$ for all $i$, since in general $S(\rho||P)\geq \sum_i q_i S(\sigma_i||P_i)$.
Furthermore, if we consider two systems with Hamiltonian $H_1$ and $H_2$, and the composite system with Hamiltonian $H=H_1+H_2$, the ergotropy is super-additive under tensor products~\cite{Francica22}, i.e., $\mathcal E(\rho_1\otimes \rho_2) \geq \mathcal E(\rho_1) + \mathcal E(\rho_2)$, thus it cannot be considered as a measure.

For $d=2$, all the passive states are completely passive states, i.e., $P=\gamma_{\beta(P)}$ with $\beta(P)=1/T(P)$ for all $P\in \Pi$, and thus $\Pi=C$. We note that $T(P|\rho)=T(P)$ does not depend on $\rho$, so that from Eq.~\eqref{eq.ergodist} we get $\mathcal E(\rho) = F(\rho)$. 
However, for $d>2$, there are free operations $\Lambda$ that do not increase the temperature for a fixed selection, i.e., $\Lambda\in \mathcal O_p$ such that $T(\Lambda(P)|\rho)\leq T(P|\rho)$, giving $T(\Lambda(P)|\Lambda(\rho)) > T(P|\rho)$.
For instance, the dephasing map $\Delta(\rho)$ preserves the passive states, so that satisfies (E,i), and does not increase the ergotropy, since the coherent contribution is nonnegative~\cite{Francica20}, i.e., $\mathcal E_c(\rho)=\mathcal E(\rho)-\mathcal E(\Delta(\rho))\geq 0$. Then, the dephasing map is a free operation, $\Delta \in \mathcal O_p$.
For the dephasing map, we get $T(\Delta(P)|\rho)=T(P|\rho)$, however we can get $T(\Delta(P)|\Delta(\rho)) > T(P|\rho)$ for $d>2$. It is enough to consider $d=3$. From an explicit calculation, for $\rho=\frac{1}{2}(\ket{\epsilon_2}\bra{\epsilon_2}+\ket{\epsilon_3}\bra{\epsilon_3}+\ket{\epsilon_2}\bra{\epsilon_3}+\ket{\epsilon_3}\bra{\epsilon_2})$, we find $P_\rho=\ket{\epsilon_1}\bra{\epsilon_1}$, $\Delta(\rho)=\frac{1}{2}(\ket{\epsilon_2}\bra{\epsilon_2}+\ket{\epsilon_3}\bra{\epsilon_3})$ and $P_{\Delta(\rho)} = \frac{1}{2}(\ket{\epsilon_1}\bra{\epsilon_1}+\ket{\epsilon_2}\bra{\epsilon_2})$, thus we find that $T(\Delta(P)|\Delta(\rho)) > T(P|\rho)$ for the passive states $P=\sum_k p_{k} \ket{\epsilon_k}\bra{\epsilon_k}$ with populations $p_{k}\geq p_{k+1}$ satisfying the inequality
\begin{equation}\label{eq.ineqex}
\ln\frac{p_1}{p_3}<\frac{\epsilon_3 -\epsilon_1}{\epsilon_3+\epsilon_2-2 \epsilon_1}\ln \frac{p_1^2}{p_2p_3}\,,
\end{equation}
which is always satisfied for $\epsilon_2=\epsilon_1$ if $p_1>p_2$.

Thus, the distance $D(\rho, \sigma)$ is defined such that, for $\sigma=P\in \Pi$ and $P\neq \ket{\epsilon_1}\bra{\epsilon_1}$, it reads
\begin{equation}
 D(\rho, P) = T(P|\rho)S(\rho||P)\,,
\end{equation}
so that the ergotropy is a based-distance monotone, since from Eq.~\eqref{eq.ergodist} it can be expressed as
\begin{equation}
\mathcal E(\rho)= \min_{P\in \Pi} D(\rho,P)\,.
\end{equation}
Then, since $\mathcal E(\rho) = F(\rho)-F(P_\rho)$, we get that the distance satisfies the property
\begin{equation}\label{eq.distanceproperty}
\min_{\gamma\in C} D(\rho,\gamma) = \min_{P\in\Pi}D(\rho, P) + \min_{\gamma\in C} D(P_\rho,\gamma)\,,
\end{equation}
i.e., the distance between $\rho$ and the set $C$ is equal to the sum of the distance between $\rho$ and the set $\Pi$ and the distance between the closest state $P_\rho$ and the set $C$.
We note that the distance $D(\rho,P)$ in general is not contractive  under the action of $\Lambda \in \mathcal O_{cp}$ such that $\eta_\Lambda(C)=1$, although it is contractive when we consider only the states $P=\gamma\in C$.
This can be proved by considering the dephasing map $\Delta\in \mathcal O_{cp}$ such that $\eta_\Delta(C)=1$ and the example previously considered to show that $T(\Delta(P)|\Delta(\rho)) > T(P|\rho)$. Similarly to Eq.~\eqref{eq.ineqex}, we find that $D(\Delta(\rho),\Delta(P))>D(\rho,P)$ if
\begin{equation}
\ln\frac{p_1}{p_3} \ln (p_2p_3)>\frac{\epsilon_3 -\epsilon_1}{\epsilon_3+\epsilon_2-2 \epsilon_1}\ln \frac{p_1^2}{p_2p_3}\ln (4p_2p_3)\,,
\end{equation}
which can be satisfied, e.g., for $\epsilon_2=\epsilon_1$, $p_1=0.6$ and $p_2=p_3=0.2$.
In general, for $\Lambda\in \mathcal O_{p}$, from Eq.~\eqref{eq.E1} we get $D(\Lambda(\rho),\Lambda(\gamma))\leq D(\rho,\gamma)$ if and only if the bound in Eq.~\eqref{eq.E2} is saturated, i.e.,
\begin{equation}
\eta_{\Lambda}(P|\rho) = \frac{S(\Lambda(\rho)||\Lambda(P))}{S(\rho||P)}\,. 
\end{equation}
The set $\mathcal O_p$ is formed by all the maps $\Lambda$ that satisfy the condition (E,ii), i.e., such that $\mathcal E(\Lambda(\rho))\leq \mathcal E(\rho)$ for all $\rho$.
In particular, maps $\Lambda\in \mathcal O_{cp}$ can belong to $\mathcal O_p$.
In general, if $\Lambda\in \mathcal O_{cp}$, from Eq.~\eqref{eq.ergoandfree}, the condition (E,ii) is satisfied, and thus  $\Lambda \in \mathcal O_p$, if and only if
\begin{equation}\label{eq.conddifF}
F(P_\rho)-F(P_{\Lambda(\rho)})\leq F(\rho)-F(\Lambda(\rho))
\end{equation}
with $F(\rho)-F(\Lambda(\rho))\geq 0$ for all $\rho$.
Then, $\Lambda \in \mathcal O_p$ if $F(P_\rho)-F(P_{\Lambda(\rho)})\leq 0$
for all $\rho$, which is difficult to satisfy, e.g., it is not satisfied if $\Lambda(\rho)\in C$ for any $\rho\notin C$ with not thermal eigenvalues, although $\Lambda\in\mathcal O_p$.
{\color{black}We get that, if $\Lambda\in \mathcal O_{cp}$, then $\Lambda \in \mathcal O_p$ only if
\begin{equation}\label{eq.condsuffdif}
\eta_\Lambda(\gamma) T(\Lambda(\gamma))+\delta_{\Lambda}(\gamma)\leq T(\gamma)\,,\forall\gamma\in C\,,
\end{equation}
where we defined
\begin{equation}
\delta_\Lambda(\gamma)= \frac{F(P_{\rho_\gamma})-F(P_{\Lambda(\rho_\gamma)})}{S(\rho_\gamma||\gamma)}
\end{equation}
and $\rho_\gamma$ such that
\begin{equation}
\max_\rho \frac{F(\Lambda(\rho))}{S(\rho||\gamma)} = \frac{F(\Lambda(\rho_\gamma))}{S(\rho_\gamma||\gamma)}\,.
\end{equation}
\begin{proof}
To prove it, it is enough to note that Eq.~\eqref{eq.conddifF} is satisfied if and only if
\begin{equation}
F(P_\rho)-F(P_{\Lambda(\rho)})\leq T(\gamma) S(\rho||\gamma)-F(\Lambda(\rho))
\end{equation}
for all $\gamma\in C$, which is equivalent to
\begin{equation}
\frac{F(P_\rho)-F(P_{\Lambda(\rho)})}{S(\rho||\gamma)}\leq T(\gamma) -  \frac{F(\Lambda(\rho))}{S(\rho||\gamma)}
\end{equation}
for all $\gamma\in C$.
By noting that $\eta_\Lambda(\gamma)$ is defined as in Eq.~\eqref{eq.etaF}, the latter condition is satisfied only if
\begin{equation}
\frac{F(P_{\rho_\gamma})-F(P_{\Lambda(\rho_\gamma)})}{S(\rho_\gamma||\gamma)}\leq T(\gamma) - \eta_\Lambda(\gamma) T(\Lambda(\gamma))
\end{equation}
for all $\gamma\in C$, which is equivalent to the condition in Eq.~\eqref{eq.condsuffdif}.
\end{proof}}
We note that the set $\Pi$ is convex, i.e., $P=\sum_i p_i P_i \in \Pi$ if $P_i\in \Pi$ for all $i$, whereas $C$ is not convex for $d>2$. 
Then, the free operations of the two sets $\mathcal O_p$ and $\mathcal O_{cp}$ can show a very different structure.
To illustrate it with an example, we consider that the ergotropy is convex, i.e., $\mathcal E(\rho)\leq \sum_i p_i \mathcal E(\rho_i)$ for any convex combination $\rho=\sum_i p_i \rho_i$.
Then, since $\mathcal E(\rho)$ is convex, any convex combination $\Lambda = \sum_i p_i \Lambda_i$, with $\Lambda_i \in \mathcal O_p$ for all $i$, belongs to $\mathcal O_p$, although we can get $\Lambda \notin \mathcal O_{cp}$ if $\Lambda_i\in \mathcal O_{cp}$ for all $i$.
\begin{proof}
From the convexity of the ergotropy we get
\begin{equation}
\mathcal E(\Lambda(\rho))\leq \sum_i p_i \mathcal E(\Lambda_i(\rho))\leq \sum_i p_i \mathcal E(\rho) = \mathcal E (\rho)\,,
\end{equation}
then (E,ii) is satisfied, so that $\Lambda\in\mathcal O_p$.
In contrast, if $\Lambda_i\in\mathcal O_{cp}$ for all $i$, since the set $C$ is not convex, given $\gamma\in C$ we can get  $\Lambda(\gamma)=\sum_i p_i \Lambda_i(\gamma)\notin C$, then (F,i) is not satisfied and $\Lambda \notin \mathcal O_{cp}$.
\end{proof}
%
In the end, we note that the condition (E,ii) is satisfied if $\Lambda$ is a unital map, i.e., $\Lambda(I)=I$, and does not increase the energy, i.e., $E(\Lambda(\rho))\leq E(\rho)$.
\begin{proof}
We note that if $\Lambda$ is a unital map, then $E(P_{\Lambda(\rho)})\geq E(P_\rho)$ (see, e.g., Ref.~\cite{Francica22}). Thus, if $E(\Lambda(\rho))\leq E(\rho)$, we get
\begin{equation}
\mathcal E (\Lambda(\rho)) = E(\Lambda(\rho))-E(P_{\Lambda(\rho)})\leq E(\rho)-E(P_{\rho})=\mathcal E(\rho)\,.
\end{equation}
\end{proof}
In particular, if $\Lambda(C)\subseteq C$, the condition $E(\Lambda(\gamma_\beta))\leq E(\gamma_\beta)$ is equivalent to $T(\Lambda(\gamma_\beta))\leq T(\gamma_\beta)$, i.e., $\eta_\Lambda(C)=1$.
If $\Lambda$ is a unitary map defined by a unitary operator $U$, so that $\Lambda (\rho)=U \rho U^\dagger$, the condition (E,ii) is equivalent to $E(\Lambda(\rho))\leq E(\rho)$. For instance the map  $\Lambda (\rho)=U_\rho \rho U_\rho^\dagger$ that extracts the ergotropy is free.
Then, there is no a work locking phenomenon~\cite{Horodecki13,Skrzypczyk14,Lostaglio15}, since all the ergotropy can be extracted with free unitary maps.

{\color{black}
\section{Monotones}\label{sec.monotones}
Let us introduce some monotones.
In the following we go beyond the quantum relative entropy by focusing on the Tsallis divergence defined as
\begin{equation}\label{eq.Tsallis}
S_\alpha(\rho||\sigma)=\frac{1}{\alpha-1}(\Tr{\rho^\alpha\sigma^{1-\alpha}}-1)\,.
\end{equation}
Similarly to the quantum relative entropy, the ergotropy can be expressed as
\begin{equation}\label{eq.monoTsa}
\mathcal E(\rho) = \min_{P\in\Pi} T_\alpha(P|\rho) S_\alpha(\rho||P)\,,
\end{equation}
which generalizes Eq.~\eqref{eq.ergodist} that is achieved in the limit $\alpha\to 1$. We defined
\begin{equation}
T_\alpha(P|\rho) = \frac{(1-\alpha)\mathcal E(\rho)}{\Tr{(P_\rho^\alpha-\rho^\alpha)P^{1-\alpha}}}\,, 
\end{equation}
so that the minimum in Eq.~\eqref{eq.monoTsa} is obtained for $P=P_\rho$.
Then we define the family of monotones
\begin{equation}
M_{p,\alpha\nu}(\rho)=\min_{P\in \Pi} T^\nu_\alpha(P|\rho)S_\alpha(\rho||P)\,,
\end{equation}
where $\nu \in [0,1]$, since we have that $M_{p,\alpha\nu}(\rho)\geq M_{p,\alpha\nu}(\Lambda(\rho))$ for $\Lambda\in \mathcal O_p$.
\begin{proof}
We aim to prove that $R_{\Lambda,\nu}(\rho)\leq 1$ for $\Lambda\in \mathcal O_p$, where we defined $R_{\Lambda,\nu}(\rho)= M_{p,\alpha\nu}(\Lambda(\rho))/M_{p,\alpha\nu}(\rho)$. We start to note that $R_{\Lambda,0}(\rho)\leq 1$ for all CPTP maps $\Lambda$ and $R_{\Lambda,1}(\rho)\leq 1$ for all  $\Lambda\in\mathcal O_p$. In order to examine the case $\nu \in (0,1)$, we note that
\begin{equation}\label{eq.deriva}
\partial_\nu M_{p,\alpha\nu}(\rho) = \frac{M_{p,\alpha\nu}(\rho)}{\nu}\ln\frac{M_{p,\alpha \nu}(\rho)}{M_{p,\alpha 0}(\rho)}\,,
\end{equation}
where $\partial_\nu M_{p,\alpha\nu}(\rho)|_{\nu=0} = M_{p,\alpha0}(\rho)\ln(M_{p,\alpha 1}(\rho)/M_{p,\alpha 0}(\rho))$.
The derivative of Eq.~\eqref{eq.deriva} can be calculated by considering the bound
\begin{equation}
M_{p,\alpha\nu}(\rho)\geq M^{\frac{\nu}{x}}_{p,\alpha x}(\rho) M^{1-\frac{\nu}{x}}_{p,\alpha 0}(\rho)\,,
\end{equation}
where $x$ is a real number, which can be easily derived by noting that
\begin{eqnarray}
\nonumber &&\min_{P\in\Pi}T^\nu_\alpha(P|\rho) S_\alpha(\rho||P) = \min_{P\in\Pi}\left(T^x_\alpha(P|\rho) S_\alpha(\rho||P)\right)^{\frac{\nu}{x}}  S^{1-\frac{\nu}{x}}_\alpha(\rho||P)\\
&&\geq  \min_{P\in\Pi}\left(T^x_\alpha(P|\rho) S_\alpha(\rho||P)\right)^{\frac{\nu}{x}} \min_{P\in\Pi} S^{1-\frac{\nu}{x}}_\alpha(\rho||P)\,.
\end{eqnarray}
By considering $\epsilon\to 0$, for $x=\nu\pm\epsilon$ we get $M_{p,\alpha\nu}(\rho)\geq M_{p,\alpha \nu\pm\epsilon}(\rho) (M_{p,\alpha \nu\pm\epsilon}(\rho)/M_{p,\alpha 0}(\rho))^{\mp\frac{\epsilon}{\nu}}= M_{p,\alpha\nu\pm\epsilon}(\rho)\mp \epsilon M_{p,\alpha\nu}(\rho)\ln(M_{p,\alpha \nu}(\rho)/M_{p,\alpha 0}(\rho))/\nu+\mathcal O(\epsilon^2) $, from which it follows Eq.~\eqref{eq.deriva}.
Thus, from Eq.~\eqref{eq.deriva} we get
\begin{equation}
\partial_\nu R_{\Lambda,\nu}(\rho) = \frac{R_{\Lambda,\nu}(\rho)}{\nu}\ln\frac{R_{\Lambda,\nu}(\rho)}{R_{\Lambda,0}(\rho)}\,,
\end{equation}
so that $\partial_\nu R_{\Lambda,\nu}(\rho)\geq 0$ when $R_{\Lambda,\nu}(\rho)\geq R_{\Lambda,0}(\rho)$, whereas $\partial_\nu R_{\Lambda,\nu}(\rho)\leq 0$ when $R_{\Lambda,\nu}(\rho)\leq R_{\Lambda,0}(\rho)$. Let us defines a general partition of $[0,1]$ such that $\partial_\nu R_{\Lambda,\nu}(\rho)\leq 0$ for $\nu\in [\nu_{2i-1},\nu_{2i}]$ and $\partial_\nu R_{\Lambda,\nu}(\rho)\geq 0$ for $\nu\in[\nu_{2i},\nu_{2i+1}]$, where $\nu_i$ are such that $0\leq \nu_i \leq \nu_{i+1} \leq 1$. For all $\Lambda \in \mathcal O_p$, for $\nu\in[\nu_{2i-1},\nu_{2i}]$ we get $R_{\Lambda,\nu}(\rho)\leq R_{\Lambda,0}(\rho)\leq 1$, whereas for $\nu\in[\nu_{2i},\nu_{2i+1}]$ we get $R_{\Lambda,\nu}(\rho)\leq R_{\Lambda,\nu_{2i+1}}(\rho)\leq 1$, where $R_{\Lambda,\nu_{2i+1}}(\rho)\leq 1$ since $\nu_{2i+1}\in [\nu_{2i+1},\nu_{2i+2}]$ or $\nu_{2i+1}=1$ and $R_{\Lambda,1}(\rho)\leq 1$. For instance, let us consider $\nu_1=0$ and $\nu_5=1$, we get $R_{\Lambda,\nu}(\rho)\leq 1$ for $\nu \in [\nu_1,\nu_2]\cup[\nu_3,\nu_4]$, $R_{\Lambda,\nu}(\rho)\leq R_{\Lambda,\nu_5}(\rho)=R_{\Lambda,1}(\rho)\leq 1$ for $\nu \in [\nu_4,\nu_5]$ and $R_{\Lambda,\nu}(\rho) \leq R_{\Lambda,\nu_3}(\rho)\leq 1$ for $\nu\in[\nu_2,\nu_3]$ since $\nu_3 \in [\nu_3,\nu_4]$ and thus $R_{\Lambda,\nu_3}(\rho)\leq 1$, then $R_{\Lambda,\nu}(\rho) \leq 1$ for all $\nu\in[0,1]$ if $\Lambda\in \mathcal O_p$.
\end{proof}
Similarly, concerning the free energy we get
\begin{equation}
F(\rho) = \min_{\gamma\in C} T_{cp,\alpha}(\gamma|\rho) S_\alpha(\rho||\gamma)\,,
\end{equation}
where we defined
\begin{equation}
T_{cp,\alpha}(\gamma|\rho) = \frac{(1-\alpha)F(\rho)}{\Tr{(\gamma_\rho^\alpha-\rho^\alpha)\gamma^{1-\alpha}}}\,. 
\end{equation}
Then we achieve the family of monotones
\begin{equation}
M_{cp,\alpha\nu}(\rho)=\min_{\gamma\in C} T^\nu_{cp,\alpha}(\gamma|\rho)S_\alpha(\rho||\gamma)\,,
\end{equation}
where $\nu \in [0,1]$, such that $M_{cp,\alpha\nu}(\rho)\geq M_{cp,\alpha\nu}(\Lambda(\rho))$ for $\Lambda\in \mathcal O_{cp}$.
In the end, we note that (v) the additivity property is never satisfied since if $\Lambda$ is a free operation of these resource theories, the operation $\Lambda\otimes \text{id}$ is not free on the Hilbert space $\mathcal H \otimes \mathcal H'$, where $\text{id}$ is the identity map on the Hilbert space $\mathcal H'$, i.e., the free operations are not completely free operations. Basically, the tensor product of two free states globally is not always a free state. Due to this tensor structure, general results derived for resource theories (see, e.g., the results in Ref.~\cite{Liu19}) cannot be applied.
As usual for these monotones, if an operation satisfies (iii) the strong monotonicity and (iv) the convexity properties then it is expected to also satisfy (ii) the monotonicity property.

}

\section{Conclusions}\label{sec.conclusions}
{\color{black} Quantum resource theories are typically useful to identify processes for the utilization of the resource for a given task. Here, we took in exam the work extractable as a possible resource.}
Work extraction in thermally isolated quantum systems can be performed through unitary cycles. The maximum work extractable is the ergotropy, which is upper bounded by the free energy. When we have many copies of the system available, the density of maximum work extractable tends to be the free energy. Thus, we defined resource theories where the free energy and the ergotropy are resources and thus monotones. Completely passive states and passive states are the  free states of the resource theories.
In particular, we discussed how, similarly to the free energy, the ergotropy is a distance-based monotone. Thus, with the aim to give a unified picture, we expressed both the free energy and the ergotropy in terms of a unique metric, with a non-contractive distance (in broad sense).
This distance is defined in terms of the quantum relative entropy and a sort of nonequilibrium temperature, which plays a certain role to characterize the free operations. In detail, any map for which the distance is contractive is a free operation for the corresponding theory. However, there are also free operations for which the distance is not contractive, finding that all the free operations can increase the temperature of the free states at most by a certain factor not smaller than 1. To summarize, thanks to the introduced distance we are able to identify the free operations based on how they act on the free states.

In conclusion, we hope that our results can be useful to understand how the work extractable can be viewed as a resource and can also find applications for the manipulation of the states of work storing centers, e.g., quantum batteries. {\color{black}Among the possible applications, we note that the charging of quantum batteries, which increases the work extractable of the batteries, can be performed only through some not free operation.}
Furthermore, we believe that our findings can also play some role in understanding what is the temperature of a nonequilibrium state.
In particular, the definition of the nonequilibrium temperature is a challenging open problem of quantum thermodynamics, e.g., a nonequilibrium temperature was recently introduced in Ref.~\cite{Alipour21}, which is different from the functional introduced by us.

\end{document}